\documentstyle[aps, psfig]{revtex}
\newcommand{\be}{\begin{equation}}
\newcommand{\ee}{\end{equation}}
\newcommand{\ba}{\begin{eqnarray}}
\newcommand{\ea}{\end{eqnarray}}
\newcommand{\bm}{\bibitem}
\newcommand{\al}{\alpha}
\newcommand{\bt}{\beta}
\newcommand{\ga}{\gamma}
\newcommand{\ep}{\epsilon}
\newcommand{\om}{\omega}
\newcommand{\lm}{\lambda}
\newcommand{\de}{\delta}
\newcommand{\De}{\Delta}
\newcommand{\Th}{\Theta}
\newcommand{\pr}{\partial}
\newcommand{\mpi}{4 m_{\pi}^2}
\newcommand{\calt}{{\cal T}^{ab}_{\mu\alpha\beta}}
\newcommand{\th}{\tanh(\beta q_0/ 2)}
\newcommand{\vq}{|\vec{q}|}
\newcommand{\la}{\langle}
\newcommand{\ra}{\rangle}
\newcommand{\usl}{u \!\!\!/}

\newcommand{\ole}{\overline}
\begin{document}

\title{Rho parameters from odd and even chirality, thermal QCD sum rules}
\author{S. Mallik and Krishnendu Mukherjee}

\address{Saha Institute of Nuclear Physics,
1/AF, Bidhannagar, Calcutta-700064, India}

\maketitle
\vspace{1cm}

\begin{abstract}
Like the even chirality correlation functions of, say, the two vector
currents, one can also consider odd chirality correlation functions to write
thermal QCD sum rules. They contain fewer non-perturbative corrections,
at least to the leading order. Here we write such a sum rule for the
correlation function of vector and tensor 'currents'. The odd and even
chirality sum rules are taken together to evaluate the effective 
parameters of the $\rho$ meson to second order in temperature.
To within errors, the results are consistent and reproduce the absence of
shift in the $\rho$ meson mass to this order.
 
\vspace{.5cm}

\end{abstract}

\section{Introduction}
\setcounter{equation}{0}
\renewcommand{\theequation}{1.\arabic{equation}}

The vacuum QCD sum rules \cite{SVZ}, when extended to finite
temperature \cite{Shap},
provide a simple means to study the properties of the QCD medium.
Below the critical temperature, they relate the temperature dependence 
of hadron parameters to the thermal average of local operators
\cite{Hatsuda,Eletsky,Mallik98}.

Generally the sum rules are written for correlation functions of two
currents, which have the same chirality, eg, of two vector or two axial
vector currents. But one may also consider correlation functions of two
quark bilinears of opposite chiralities \cite{Mallik82}. Then their
operator product expansions will be dominated, in each dimension,
by the operators of odd chirality, the contributions of even chirality ones
being highly suppressed by a factor of (small) quark mass  
in their coefficient functions. As a result the leading nonperturbative
corrections will be given by a single operator, namely, the quark
condensate; new odd chirality operators appear only as non-leading
terms.
 
Here we write down the sum rule for the correlation function
of the vector current and the tensor 'current'. Then both the present
vector-tensor sum rule and the two vector-vector sum rules derived earlier get
contributions from the same set of intermediate states with 
$J^{PG} =1^{-+}$. We consider the subtracted sum rules, obtained
by subtracting out the vacuum sum rules from the corresponding thermal ones.
The only intermediate states with significant contributions are then the 
non-resonant $2\pi$ state and the $\rho$-resonance. The three sum rules
form a convenient set for evaluation \cite{Comment1}.

We confine our evaluation at low temperature $T$ 
to leading order, which is $T^2$
in the chiral limit. Generally the thermal sum rules are complicated by the
presence of Lorentz non-scalar operators, in addition to the Lorentz scalars
contributing to the vacuum sum rules \cite{Shur}.
But to order $T^2$ non-scalar
operators cannot contribute to these sum rules. 
Thus to this order, the contributing operators
are the same as at zero temperature, their vacuum expectation values being
replaced by the thermal averages.

In Sec II we describe the kinematics and present the result of operator
product expansion. In Sec III we saturate the sum rules with $2\pi$ and
$\rho$ states. In Sec IV the sum rules are evaluated for the 
parameters of the $\rho$ meson 
to order $T^2$ in the chiral
limit. Finally in Sec V we discuss the different aspects of the thermal sum
rules and compare our evaluation with earlier ones.

\section{Sum Rules}
\setcounter{equation}{0}
\renewcommand{\theequation}{2.\arabic{equation}}

We restrict here to the better known non-strange channels of unit isospin. The
quark bilinears in this channel are,
\[S^a(x), V^a_{\mu}(x), T^a_{\mu\nu}(x), A^a_\mu(x), P^a(x) =
\bar{q}(x)(1, \ga_{\mu}, \sigma_{\mu\nu}, \ga_\mu\ga_5, \ga_5){\tau^a\over
2}q(x),\]
$q(x)$ being the field of the $u$ and $d$ quark doublet and $\tau^a$ the
Pauli matrices.
Note that $S^a(x)$, $P^a(x)$ and $T^a_{\mu\nu}(x)$ are odd and 
$V^a_{\mu}(x)$ and $A^a_{\mu}(x)$ are
even under $\ga_5$ transformation on the quark field.

As an important example of odd chirality correlation functions, consider the
thermal average of the time ordered ($T$) product of the vector 
current and the tensor 'current',
\be
\calt (q) = i \int{d^4x} e^{iq\cdot x}
\left \langle T\left(V^a_{\mu}(x)
T^b_{\al\bt}({\rm o})\right) \right \rangle.
\ee
Here the thermal average of an operator $O$ is denoted by 
$ \la O \ra$,
\[\langle O\rangle =Tr e^{-\beta H} O/Tr e^{-\beta H} ,\]
where $H$ is the QCD Hamiltonian, $\beta$ is
the inverse of the temperature $T$ and $Tr$ denotes the trace over any complete
set of states. 

As usual, it is convenient for kinematics to restore Lorentz
invariance by introducing the four-velocity $u_{\mu}$ of matter. Then
we can build the Lorentz scalars, $\om = u\cdot q$ and 
$\bar{q}=\sqrt{\om^2-q^2}$, representing the time and the space components of
$q_{\mu}$ in the matter rest frame, $u_{\mu}=(1,0,0,0)$ \cite{Weldon82}. 
We now choose the three independent kinematic covariants as,
\ba
P_{\mu\al\bt} &=& \eta_{\mu\al} q_{\bt} -\eta_{\mu\bt} q_{\al} 
, \nonumber \\ 
Q_{\mu\al\bt} &=&  q_{\mu}(q_\alpha u_\beta - q_\beta u_\alpha) 
- q^2(\eta_{\mu\alpha}u_\beta - \eta_{\mu\beta}u_\alpha),\nonumber\\
R_{\mu\al\bt} &=&  u_{\mu}(q_\alpha u_\beta - q_\beta u_\alpha) 
- \om(\eta_{\mu\alpha}u_\beta - \eta_{\mu\beta}u_\alpha).
\ea
The kinematic decomposition now reads,
\be
\calt (q)=i\de^{ab} (P_{\mu\al\bt} T_1 + Q_{\mu\al\bt} T_2  
+ R_{\mu\al\bt} T_3), 
\ee
where the invariant amplitudes $T_{1,2,3}$ are functions of $\om$ and $q^2$.
In all computations, however, we shall revert back to the matter rest
frame.

Only the amplitude $T_1$ survives at zero temperature. As we shall see below,
at finite temperature the leading contributions (to order $T^2$) are also
contained in $T_1$. So, working to the leading order, we have only to write the
sum rule for this amplitude.

The advantage with odd chirality correlation functions becomes evident
from an enumeration of local operators. The unit operator corresponds to the
perturbative result. The non-perturbative power corrections begin with
operators of dimension three and four. At dimension three, we have 
$\bar{q} q$ and $\bar{q} \usl q$ 
belonging to odd and even chirality respectively.
(The operator $\bar{q} \usl q$ actually cannot contribute for zero chemical
potential.) At dimension four, we have $trG_{\mu\nu} G^{\mu\nu}, \;
u^{\mu} \Th_{\mu\nu}^f u^{\nu}$ and $u^{\mu} \Th_{\mu\nu}^g u^{\nu}$.
Here the gauge field strength $G_{\mu\nu}=g{\lambda^a\over 2}G^a_{\mu\nu}$, 
$\lambda^a$ ($a=1, \cdots 8$) are the $SU(3)$ Gell-Mann matrices 
and $g$ is the QCD coupling constant. The operators 
$\Th_{\mu\nu}^{f,g}$ are the energy momentum tensors for the quarks and
gluons. (Note that in the matter rest frame, the latter two are
just the energy densities. Also the operator $\bar{q} u\cdot D q$ where
$D_{\mu}$ is the covariant derivative, is of odd chirality, but it can be
reduced to $\hat{m}\bar{q} \usl q$ by using the equation of motion for the quark
field.)  All these operators of dimension four are of even chirality.
Thus up to dimension four, only the operator $\bar{q}q$ can contribute
significantly to the power correction for an odd chirality correlation
function. At dimension five, there is only one Lorentz scalar operator,
$\bar{q} \sigma^{\mu\nu}G_{\mu\nu}q$. In addition there are several Lorentz
non-scalar operators contributing at finite temperature\cite{Jin}.   

There is a simple configuration space 
approach\cite{Fritzsch} to the operator
product expansion giving the Wilson coefficients of all operators, both
scalars and non-scalars\cite{Hub,Mallik97}. Using this method and
restricting to scalar operators,  we get for the product under consideration,
\be 
T V^a_{\mu}(x)T^b_{\al\bt}(0) \rightarrow \de^{ab}(\eta_{\al\mu}
x_\bt - \eta_{\bt\mu} x_\al)\Biggl\{{3\hat{m}\over 2\pi^4}
{1\over (x^2-i\ep)^3}{\bf 1}
+{1\over 2\pi^2} {1\over (x^2 -i\ep)^2} \bar{u}u 
+ {1\over 24\pi^2}{1\over (x^2 - i\ep)} O_5
+ \cdots\Biggr\},
\ee
where $O_5$ is the dimension five operator, 
$O_5=\bar{u}\sigma_{\mu\nu}G^{\mu\nu}u$,
and the dots represent operators of still higher dimensions. 
We have assumed $SU(2)$ flavour symmetry; $\hat{m}$ is 
the degenerate mass of $u$ and
$d$ quarks and $\bar{u}u =\bar{d}d={1\over 2}\bar{q}q.$
The Fourier transform gives for space-like momenta $(Q^2=-q^2\geq 0)$,
\be
\calt (q) \rightarrow i\de^{ab} (\eta_{\mu\al} q_{\bt} -\eta_{\mu\bt} q_{\al})
\left \{ -{3\hat{m}\over {8\pi^2}}log (Q^2/{\mu^2}) -{1\over Q^2}
\la \bar{u}u\ra + {1\over 3(Q^2)^2}\la O_5 \ra
+ \cdots \right \}.
\ee
Here $\mu$ $(\simeq 1$ GeV ) is the renormalisation scale.

We wish to include the renormalization effects on the operators 
$T^a_{\mu\nu}$ and $\bar{u}u$. 
The coefficient $C(Q^2/{\mu^2}, g(\mu))$ of $\bar{u} u$ satisfies
 \cite{Peskin},
\be
\left ( \mu {\pr\over {\pr \mu}} + \bt (g){\pr\over {\pr g}}
+\ga_1 -\ga_2 \right ) C(Q^2/{\mu^2}, g(\mu)) =0, 
\ee
where $\bt (g)=-b {g^3\over {(4\pi)^2}}, \; b=9$. The anomalous dimensions
$\ga_1$ and $\ga_2$ of the operators $T^a_{\mu\nu}$ and $\bar{u} u$ 
can be easily calculated to give
$\ga_1 =d_1 {g^2\over {(4\pi)^2}}$ and $\ga_2 =d_2  {g^2\over {(4\pi)^2}}$
with $d_1 =8/3$ and $d_2 =-8$.
The solution to (2.6) may be written as
\[  C(Q^2/{\mu^2}, g(\mu)) = a(Q^2) C(1, \bar{g} (Q^2)),\]
where $C(1, \bar{g} (Q^2))$ is the lowest order result obtained above and
\[ a(Q^2)= \left ( {log(Q^2/\Lambda^2)\over {log (\mu^2/\Lambda^2)}} \right )
^{(d_1 -d_2)/{2b}} .\]
The strong interaction scale is $\Lambda \simeq 200$ MeV.

The other element needed for the sum rule is the spectral representation
for the correlation function. The $T$ product at finite temperature has the
so-called Landau representation in the variable $q_0$ at fixed $\vq$
 \cite{Landau}. At
points on the imaginary axis $(q_0^2=-Q_0^2, Q_0^2>0)$, the representation
for the invariant amplitudes are, up to subtractions, given by
\be
T_{1,2,3}(q_0^2,\vq) =
\int^{\infty}_0dq^{\prime^2}_0 {N_{1,2,3}(q^{\prime}_0,\vq)\over
{q^{\prime^2}_0 + Q_0^2}}.
\ee
where
\[N_{1,2,3} (q_0,\vq)={\pi}^{-1} Im T_{1,2,3} (q_0^2,\vq) \th .\]

\section{Saturation}
\setcounter{equation}{0}
\renewcommand{\theequation}{3.\arabic{equation}}

In the channel under consideration, the $\rho$-meson  
dominates the absorptive part. To find this contribution we write the
matrix elements of the currents,
\be
\la 0|V_{\mu}^a|\rho^b(k)\ra = \de^{ab} F_{\rho} m_{\rho} \epsilon_{\mu},\qquad
\la0|T_{\al\bt}^a|\rho^b(k)\ra =i \de^{ab} G_{\rho} ( \epsilon_{\al} k_{\bt}-
\epsilon_{\bt} k_{\al}).
\ee
Here $F_{\rho}$ and $G_{\rho}$ are two coupling constants and
$m_{\rho}$ and $\epsilon_{\al}$ are the mass and the polarization
vector of the $\rho$-meson. $F_{\rho}$ is measured by the electronic decay
rate of $\rho^0$, $F_{\rho} =153$ MeV \cite{Particles}. Though the value of
$G_{\rho}$ is not available directly from experiment, it can be obtained
from one of the U(6)
symmetry relations for the wave functions of a quark-anti-quark
pair \cite{Leutwyler74}. Defining the pion decay constant $F_{\pi}$ by
\be
\la 0| A_{\mu}^a |\pi^b (k)\ra =i\de^{ab} k_{\mu} F_{\pi},
\ee
this relation is
\be
G_{\rho}={1\over 2}(F_{\rho} +F_{\pi}). 
\ee
As $F_{\pi}=93$ MeV, we get $G_{\rho}=123 $ MeV.
This value of $G_{\rho}$ is also obtained from the QCD sum rules for the
correlation function of two tensor 'currents' \cite{Mallik82}.
Comparing (3.1) with the matrix element of the 
$\rho$-meson field operator $\rho_{\mu}^a(x)$,
$ \la0|\rho_{\mu}^a|\rho^b\ra = \de^{ab} \epsilon_{\mu}$,
we get the operator relations, 
\[V_{\mu}^a (x)=m_{\rho} F_{\rho}\rho_{\mu}^a (x),\qquad
T_{\al\bt}^a (x)=G_{\rho}(\partial_{\al}\rho_{\bt}^a
-\partial_{\bt}\rho_{\al}^a).  \]

The absorptive parts are now given essentially by that of the $\rho$-meson
propagator at finite temperature. Working in the real time formulation of
the thermal field theory \cite{Niemi}, 
it is the $11$-component of the $2\otimes 2$ matrix propagator.
We get
\be
N_1 (q^2)= m_{\rho} F_{\rho} G_{\rho}\de (q^2-m_{\rho}^2),\qquad
N_2(q^2)=N_3(q^2)=0 .
\ee
This calculation must be interpreted as one in the effective field theory at
finite temperature, where loop corrections make the
parameters $m_{\rho}, F_{\rho}$ and $G_{\rho}$ temperature dependent.
(At finite temperature, each of the particle-current coupling constants 
in (3.1-2) has, in  general, different temperature dependence for
the time and space components of the currents \cite{Pisarski}.
But this bifurcation takes place only in orders higher than $T^2$.
So it does not concern us here.) 

At lower energies there is the contribution of non-resonant two-pion
state. Although small compared to the $\rho$-meson contribution, it
describes the interaction of the currents with the pions in the heat bath and
may assume importance in the difference sum rules we shall consider below.
We find this absorptive part by writing the field theoretic expression for
the pion loop at finite temperature. The pionic content $J_{\mu}^a(x)$ 
of the quark vector current $V_{\mu}^a(x)$ is given, to lowest order, by
\[ V_{\mu}^a(x)\rightarrow J_{\mu}^a(x) =\epsilon^{abc} \phi^b(x) \partial_{\mu}
\phi^c(x).\]
However the pionic content $S_{\al\bt}^a(x)$ of the quark tensor 'current'
$T_{\al\bt}^a(x)$ is not immediately known, as it is not a symmetry current. From
its index structure it may be written as
\[T_{\al\bt}^a(x)\rightarrow S_{\al\bt}^a
(x)=c\epsilon^{abc}\partial_{\al}\phi^b (x)\partial_{\bt}\phi^c (x),\]
where we determine the constant $c$ by comparing the two
divergences,
\[ \partial^{\al} T_{\al\bt}^a(x) =2\hat{m} V_{\bt}^a(x) +\cdots,\qquad
\partial^{\al}S_{\al\bt}^a(x) = c m_{\pi}^2 J_{\bt}^a(x) +\cdots,\]
the dots standing for higher derivative terms. Like $V^a_{\mu}(x)$ and
$J^a_{\mu}(x)$, we can also identify $ T_{\al\bt}^a(x)$ and $S_{\al\bt}^a(x)$, 
giving $c=2\hat{m}/m_{\pi}^2=- F^2_\pi/\la 0\mid\bar{u}u\mid 0\ra$, on
using the Gell-Mann, Oakes and Renner relation \cite{Gell-Mann}.

With the pionic version of $V_{\mu}^a(x)$ and $T_{\mu\nu}^a(x)$, it is simple to
evaluate (2.1) to lowest order as,
\be
\calt (q)= -c \de^{ab}\int{d^4k\over (2\pi)^4}(2k-q)_{\mu}
(q_{\al}k_{\bt}-q_{\bt}k_{\al})
\De_{11}(k)\De_{11}(k-q),
\ee
where $\De_{11}$ is the $11$- component of the thermal pion
propagator. Its absorptive part can be calculated in the same way as for
the vector-vector correlation function \cite{Mallik98}. 
As we need to write the sum rule for the amplitude $T_1$, we quote
the results for this amplitude only. In the time-like
region (superscript +),  
\be
N_1^+ =
{c q^2\over 128\pi^2}\int^{v}_{-v} dx
(v^2(q^2)-x^2)\{1+2n((\vq x+q_0)/2)\},
~~~~{\rm for}~~~ q^2\geq 4m_{\pi}^2,
\ee
while in the space-like region (superscript $-$),
\be
N_1^- =
{c q^2\over 128\pi^2}\int^{\infty}_v dx
(v^2(q^2)-x^2)
\{n((\vq x-q_0)/2)-n((\vq x+q_0)/2)\},
~~~~{\rm for}~~~ q^2\leq 0
\ee  
Here we have defined the functions $v(z)=\sqrt{1-4m_{\pi}^2/z}$ and
$n(z)=(e^{\bt z} -1)^{-1}$.

For$\vq\rightarrow 0$, they reduce to simple expressions. In the time-like
region,
\be
N_1^+={c q^2_0v^3(q_0^2)\over {96\pi^2}} \{1+2n(q_0/2)\}.
\ee
In the space-like region, $q_o^2=\lm \vq^2$, with $0\leq \lm \leq 1$. 
Thus in the
limit $\vq \rightarrow 0$, $N^-_1$ and its contribution to the
spectral representation (2.7) are zero. 

We can now write a spectral sum rule by equating at a space-like point 
the operator product expansion (2.5) to the spectral
representation (2.7) with $\rho$ and $2\pi$-contributions. Taking Borel
transform as usual 
to get rid of any subtraction constant and to improve convergence
of the spectral integral, we arrive at the 
thermal QCD sum rule,
\ba
 & & m_{\rho} (T) F_{\rho}(T) G_{\rho} (T) e^{-m_{\rho}^2(T)/M^2} - 
{F_\pi^2\over 96\pi^2\la 0\mid\bar{u}u\mid 0\ra} 
\int_{\mpi}^\infty ds s e^{-s/M^2} v^3(s)(1+2n(\sqrt{s}/2))\nonumber\\
&=&  {3\hat{m}\over 8\pi^2}M^2 - a(M^2)\la\bar{u}u\ra 
+ {1\over 3M^2}\la O_5 \ra.
\ea

\section{Evaluation}
\setcounter{equation}{0}
\renewcommand{\theequation}{4.\arabic{equation}}

As already stated, we shall evaluate sum rules to order
$T^2$ in the chiral limit. Since the continuum integral is $0(T^4)$, 
the odd chirality sum rule (3.9), after subtraction of the corresponding 
vacuum sum rule, simplifies to
\be
m_\rho(T) F_\rho(T) G_\rho(T) e^{- m_\rho^2(T)/M^2}
- m_\rho F_\rho G_\rho e^{- m_\rho^2/M^2}
= - a(M^2)\ole{\la\bar{u}u\ra} + {1\over 3M^2}\ole{\la O_5 \ra}.
\ee
Here the bar over the thermal average indicates subtraction of the 
corresponding vacuum expectation value.

For the sake of consistent evaluation in a closed framework, we
augment this sum rule with the two even chirality sum rules for
the vector-vector correlation function derived earlier \cite{Mallik98}. 
In the chiral limit and omitting terms of order higher than $T^2$, they
become
\ba
F_\rho^2(T) e^{-m_\rho^2(T)/M^2} - F_\rho^2 e^{-m_\rho^2/M^2}
+ I_T(M^2) &=& -{\pi\over 4}{\ole{\la O_6\ra}\over M^4},\\ 
F_\rho^2(T) m_\rho^2(T) e^{-m_\rho^2(T)/M^2} - 
F_\rho^2 m_\rho^2 e^{-m_\rho^2/M^2}
 &=& {\pi\over 2} {{\ole{\la O_6\ra}}\over M^2},
\ea
where
\be
I_T(M^2) = {1\over 24\pi^2}\int_0^\infty ds (1 + e^{-s/M^2})
{1\over e^{\sqrt{s}/2T}-1} = {T^2\over 9} + 0(T^4)
\ee
and $O_6$ is the four-quark operator, 
$O_6 = \alpha_s \bar{q}\gamma_\mu\gamma_5\lambda^a\tau^3q~  
\bar{q}\gamma^\mu\gamma_5\lambda^a\tau^3q$ \cite{Comment2}.
This operator was ignored in \cite{Mallik98}.

At low temperature the heat bath consists primarily of dilute pion gas. The
thermal trace can then be approximated by the vacuum and the one pion state.
The pion matrix element of the operator can be evaluated by using PCAC and
soft pion technique \cite {Hatsuda}. One gets,
\be
\la\bar{u}u\ra = \la 0\mid\bar{u}u\mid 0\ra\left(1 - {T^2\over 8F_\pi^2}
\right),~~ 
\la O_5\ra = \la 0\mid O_5\mid 0\ra\left(1 - {T^2\over 8F_\pi^2}
\right),~~ 
\la O_6\ra = \la 0\mid O_6\mid 0\ra\left(1 - {T^2\over 3F_\pi^2}\right). 
\ee
where the vacuum expectation values are all known 
\cite{Gasser,Belyaev,Comment3},
\be
\la 0\mid\bar{u}u\mid 0\ra = -(225 {\rm MeV})^3,~~ 
\la 0\mid O_5\mid 0\ra = m_o^2 \la 0\mid\bar{u}u\mid 0\ra,~~ 
\la 0\mid O_6\mid 0\ra = 6.5\times 10^{-4} {\rm GeV}^6. 
\ee
with $m_o^2=1.0{\rm GeV}^2$. 
If we write 
\be
m_\rho(T)=m_\rho\left(1+a{T^2\over F^2_\rho}\right),~~
F_\rho(T)=F_\rho\left(1+b{T^2\over F^2_\rho}\right),~~
G_\rho(T)=G_\rho\left(1+c{T^2\over F^2_\rho}\right),
\ee
the sum rules (4.1-3) predict the values of the constants
$a$, $b$ and $c$ as
\ba
a &=& e^{m_\rho^2/M^2}\Biggl\{{1\over 18} - {K_2\over 4M^4}
\biggl(1 + {2M^2\over m_\rho^2}\biggr)\biggr\} ,\\ 
b &=& -e^{m_\rho^2/M^2}\Biggl\{{1\over 18}
\biggl(1 - {m_\rho^2\over M^2}\biggr) + {K_2\over 4M^4}
\biggl(1 + {m_\rho^2\over M^2}\biggr)\biggr\} ,\\
c &=& e^{m_\rho^2/M^2}\Biggl\{{m_{\rho}^2\over 18M^2}
 + K_1 \biggl(a(M^2) - {m_0^2\over 3M^2}\biggl)
- {K_2\over 4M^4}\biggl({m_\rho^2\over M^2}
- {2M^2\over m_\rho^2} + 2\biggr)\Biggr\}, 
\ea
where
\[K_1={\la 0\mid\bar{u}u\mid 0\ra\over 4m_\rho F^2_\pi(1 + F_\pi/F_\rho)},
~~~K_2= {\pi\over 6F_\pi^2}\la 0\mid O_6\mid 0\ra .\]

It should be noted that the parameters $m_\rho$, $F_\rho$ and
$F_\pi$ in eqn.(4.8-10) refer to their chiral limits and not
the physical values. There does not appear to exist any  calculation of
$F_\rho$ to this limit in the literature \cite{Bijnens}. Under
this circumstance, let us evaluate $a$, $b$ and $c$ with
the physical values of these parameters as quoted in Sec. III and hope
that the results will not be affected much. 


\vspace{0.8cm}

\begin{center}
\begin{tabular}{|c|c|c|c|}
\hline
$M^2$ & $a$ & $b$ & $c$ \\
(${\rm GeV}^2$) & & & \\
\hline
~~~~0.8~~~~ & ~~~~0 ~~~~& ~~~~-0.086~~~~ & ~~~~-0.21~~~~ \\ 
 & & & \\\hline
1.0 & 0.020 & -0.069 & -0.24 \\ 
 & & & \\\hline
1.2 & 0.035 & -0.063 & -0..26 \\ 
 & & & \\\hline
2.0 & 0.049 & -0.057 & -0.30 \\ 
 & & & \\\hline
4.0 & 0.054 & -0.056 & -0.33 \\ 
 & & & \\\hline

\end{tabular}
\vspace{0.3cm}

Table 1: Coefficients of $T^2$ in $m_\rho(T)$,
$F_\rho(T)$ and $G_\rho(T)$ at different values of $M^2$.
\end{center}

\vspace{0.5cm}

Table I shows the evaluation of 
the constants $a$, $b$ and $c$ for different values 
of the Borel parameter $M^2$ over the range $0.8\leq M^2\leq 4$ in 
${\rm GeV}^2$. It is seen that as $M^2$ increases, the values change but
rather slowly. Observe that at $M^2=0.8 GeV^2$,  
one gets $a=0$, reproducing the result that the hadron masses do not shift
to order $T^2$ \cite{Leutwyler90,Eletsky93}. 
This suggests that we take the central values of all the
constants at this value of $M^2$ in the prediction of our sum rules,
\be
a=0, \qquad  b=-.086, \qquad c=-.21,
\ee
with an error of about 25\%, estimated from their variations in the range of
$M^2$ considered.

Eq(3.3 ) relating $G_{\rho}$ to $F_{\rho}$ and $F_{\pi}$ is used above only at
zero temperature. Being a symmetry relation one expects it to be valid also
at finite temperature,
\be
G_{\rho}(T)={1\over 2}(F_{\rho}(T)+F_{\pi}(T)).
\ee
This equation provides another determination of $c$, if we take $F_{\rho}(T)$ as
determined above and 
\be
F_{\pi}(T)=F_{\pi} \left(1-{T^2\over {12 F_{\pi}^2}}\right)
\ee
as obtained from chiral perturbation theory. It gives $ c=-.14$, in fair
agreement with $c=-.21$ obtained above.       

\section{Discussion}
\setcounter{equation}{0}
\renewcommand{\theequation}{4.\arabic{equation}} 

We derive here an odd chirality thermal QCD sum rule. More such sum rules
can be written by considering other correlation functions of odd chirality,
for example, of the axial vector and pseudo-scalar currents. Generally they
bring in a different set of operator expectation values, but are fewer in
number compared to those in the even chirality sum rules, at least to the
leading order. It would be useful to consider both sets of sum rules for
numerical evaluation.

The behaviour of the absorptive parts in the space-like region in different
sum rules is worth mentioning here. Such absorptive parts are a
characteristic feature of thermal field theories and here it describes pion
absorption by the current from the heat bath and its emission into it. For
nonzero $\vq$, it is present in all the sum rules. However,
as $\vq \rightarrow 0$, it does or does not survive depending on the
kinematics. In the sum rules considered here, it survives only in the sum
rule (4.2) , obtained from the longitudinal amplitude defined in Ref.
\cite{Mallik98}.

We have evaluated the sum rules to order $T^2$ in the chiral limit. 
The sum rules are expected to be 
well saturated to this order and should be independent of
the Borel parameter $M^2$ over a larger range. The subtraction eliminates
the continuum contribution beyond about $1 $GeV, it being practically 
independent
of $T$ for $T< 150 MeV$, say. There are thus no other significant source of
contribution in the spectral representation besides those already included,
namely, the $2\pi$ -continuum and the $\rho$-resonance.

As regards the 
operators, only those which are Lorentz scalars and appear already
in the corresponding vacuum sum rules, may produce
terms of order $T^2$. We note, however, that while the $T^2$ contributions
in the spectral representation arise {\it only} from $2\pi$ and $\rho$ states,
scalar operators of {\it all} dimensions can 
in general contribute to this order,
though, of course, the usual Borel suppression operates for operators of
higher dimensions.
 
The other point to note is that the perturbative contribution,
which is the coefficient of the unit operator, is eliminated by subtraction.
So the higher dimension operators, which constituted only (power)
corrections to the perturbative result in the vacuum sum rules, 
are now assuming the leading role.
Thus the subtracted version of sum rules considered here brings out clearly 
the sensitivity of the thermal sum rules to any uncertainty in the 
operator expectation values. 

The present work is a continuation of our earlier work \cite{Mallik98}. 
There our main concern was to include all the operators up to dimension 
four with their renormalisation effects in the vector-vector sum rules. But
the omission of the scalar, four quark operator was
not justified. For, while the dimension four operators are significant at
higher temperatures near the critical point, at low temperatures it is only
this dimension six operator which dominates, contributing to order $T^2$
in the chiral limit.
 
The most
extensive earlier work is that of Hatsuda et al \cite{Hatsuda}. They include
contributions up to order $T^6$ from the operators and plot the $T$-dependence
of $m_{\rho}(T)$ and $F_{\rho}(T)$ ( and also parameters of other
resonances) but do not obtain the coefficients of $T^2, T^4$ etc.
(Incidentally their evaluation up to order $T^6$ is incomplete, because
they include contributions up to order $T^4$ and $T^6$ only when they are
leading in some operators, but neglect non-leading contributions even 
to order $T^4$ in others, eg, $\la \bar{u} u\ra$.)
Eletsky et al \cite{Eletsky95} calculate the $\rho$ meson mass to order
$T^4$, but do not take the mixing of dimension four, Lorentz non-scalar
operators under renormalisation.

\end{document}